\documentclass{cta-author}
\usepackage[export]{adjustbox}

\begin{document}


\title{Inferring demand from partially observed data to address the mismatch between demand and supply of taxis in the presence of rain}

\author{\au{Seyyed Yousef Oleyaei-Motlagh $^{1\corr}$}, \au{Adan Ernesto Vela$^{1}$}}

\address{\add{1}{Dept. of IEMS, University of Central Florida,
	Orlando, Florida, United States}
\email{seyyed@knights.ucf.edu}}

\begin{abstract}
 Analyzing mismatch in supply and demand of taxis is an important effort to understand passengers' demand. In this paper, we have analyzed the effect of rain on the demand for yellow taxis in city-wide as well as in a point of interest in New York City. Because a pickup event is a realized demand, we studied empty travel time, the number of pickups per driver, the average amount of income per drive indices to infer demand from taxis data of 2013. Findings highlight that the higher demand exists because of many short-trips during the rain. This paper illustrates the change in passengers' demand increased by the onset of weather condition.
\end{abstract}

\maketitle

\section{Introduction}\label{sec1}

\setlength\parindent{24pt}{In today's connected transportation systems, different sensor readings from different sources e.g., wireless networks, satellites, and observatory weather stations are easily accessible for researchers to infer demand for their transportation mode of interest. In large cities, taxis provide point-to-point traveling service that are among the most convenient transportation modes of choices to accommodate customers special needs. It is widely reported that some passengers experience long-wait time for taxis in rainy weather while transit users experienced higher humidity that make the alternative inconvenient mode of transportation during summers in NYC. The characteristics of weather are the key to unlock the taxi demand problem in different weather conditions. Thus, demand for a fleet of taxi vehicles without prearrangement, is driven by myriad of factors including special events, time of day (TOD), weather (rain, temperature, humidity, ...), presence of empty vehicle in the line of sight of passenger and its recognition of passenger, and etc. Given multitudinous of factors, we can only infer demand in the settings of time series of variable of interest.\\}\par
\setlength\parindent{24pt}{Prior to emergence of ride-sharing technologies, the yellow taxi cabs were ubiquitous in New York City (NYC). In 2013 the yellow taxi cabs were the only sanctioned vehicle to pick up street hail passengers in Manhattan, NYC. In recent years, taxi business has been challenged  by introducing ride-sharing and autonomous vehicles. Regardless of technologies we were can transfer the knowledge from outstanding change in e-hails demand due to weather to monopoly of medallions. We studied mismatch between supply and realized demand, which is the number of met pickups. Studying the demand for a particular cluster of taxi transportation, utilizing historical pickups and weather information is an important effort to socially understand the demand  as a function of input variables of location, time of day, and weather condition e.g., rain, wind, high-temperature, and snow. Therefore, we adopted best practices of transportation including ride-sharing to address recent changes \cite{Tao2018_TRPC_WEATHER_QUETSTION}.\\}\par
\setlength\parindent{24pt}{\setlength\parindent{24pt}{To address these rapid changes, most recently, Canadian Urban Transit Association (CUTA) \cite{miller2018canadian_CUTA} surveyed the factors affecting ridership at the aggregate level including weather, ride sharing, and ridership for adults and students to find the trend in transit ridership \cite{gaudry1996demand,Gaudry1974TranspResearch,miller2018canadian_CUTA}. In a study \cite{Arana_2014_Part_A_Trsp_Research_Influence_of_weather} the impact of meteorological conditions on the number of public bus trips used for different purposes has been studied in Gipuzkoa, Spain. In another study \cite{chakour_examining_2016_Journal_of_Transport_Geography} researchers found that the higher demand is because of the high number of bus lines, bus stops, and the transfer points between a bus terminal or a metro station in that area that attract many customers. Understanding the taxi demand in different weather conditions requires fine-grained data-sets. Such data can reveal important characteristics of taxi passenger travel behavior in different situations. Such analysis are intensively dependent on time and environmental factors, and the state of system. Meantime, visualization is a great tool to describe complex factors in easily understandable framework. The time series data and visualization were obtained using various database technologies including MySQL, Pandas and DASK libraries of Python along with applying appropriate aggregation on historical data sets. This paper analyzes taxis supply and passenger demand using data for medallions (yellow taxis) in 2013 in New York City (NYC).\\}\par
\setlength\parindent{24pt}{\setlength\parindent{24pt}{The main objective of this paper was to present innovative use of measures on historical pickups and supply of the yellow taxis in different zones of NYC, at different times and at different weather conditions to infer the demand. 
This paper is organized as follows. In the related works section, we provide a background for the factors affecting taxi supply and passenger demand. In the data section, we describe weather data and provide context for the analysis in the next section. In details in section IV, we describe methods for transforming data. In the result, we describe time series of variables of interests and present statistical analysis. In the conclusion, we compare our results. Finally, based on conclusions and limitations of current study, future works are suggested.\\}\par
\section{Related Work}\label{sec2}
\setlength\parindent{24pt}{Previous research has identified that the bad weather influences on taxi-drivers income \cite{kamga2013_TRB92, Kamga2015_TPT}. Based on 147 million records of yellow taxi pick-ups in New York City in 10 months of 2010, researchers have obtained different income level for different time of the day, the day of week (Mon, Tue, etc), and the month of year \cite{kamga2013_TRB92, Kamga2015_TPT}. A high-resolution pick-up prediction model is based on high-resolution GPS data in the current study.\\}\par
\setlength\parindent{24pt}{Taxi drivers either own a medallion or they could lease one to work in one of two shifts (mostly two but three would be a possibility) determined in the lease agreement. Flexibility and on-time service could let passengers hail taxis from every part of NYC in different time and weather conditions. However, it was less likely to hail an unoccupied cab on a rainy day in Manhattan in the preridesharing era \cite{Kamga2015_TPT}\cite{KyungLee_2017TPT_Taylor}. Many taxi drivers prefer to work in evenings shift when they look for higher profit and lease the car for other shifts. In fact, one of the factors affecting leasing price is the time of the start of shift. The dispersion of yellow taxis in different regions (taxi supply) is governed by the dynamics of taxi-passenger flow and spatiotemporal demand in absence of a central dispatch system in New York City. 
	
\setlength\parindent{24pt}{The taxis' ridership differs from transit. The largest body of research, which explains the variations in ridership relative to the weather belongs to the public transportation dates as far back as the 1980s; while, there is no study to address the impact of weather on taxi companies in the literature \cite{Bertness1980JAM,HofmannITSC2005}. In \cite{Bertness1980JAM} the author has found that the summer rainfall, decreased ridership for mass transit approximately 3-5\%. Because there are differences between the public taxi and transit ridership; thereafter, effects of weather on taxis might be different, gesturing a new research direction in transportation literature.\\}\par
\setlength\parindent{24pt}{Even though, some geographical locations have captured higher pick-ups rate than other regions the pick-ups rate within the region is dynamic. The pick-up pattern depends on time varies across NYC regions. For example, in residential areas outside of Manhattan, more pick-ups occur in the mornings; however, in business-dominant areas in Manhattan there is a higher rate of pick-ups in the mid-nights. Taxi-trips are short distance and duration mode of transportation \cite{KyungLee_2017TPT_Taylor}. Every pick-ups ends up in a destination that is close to its origin, except airport pick-ups, which have longer trip distance and duration e.g., taxi pick-ups at JFK airport, which has a policy in-action of fixed-route/price to Manhattan\footnote{TLC reports that in 2014, 20\% of all the trip distance were less than a mile (about 20 Manhattan blocks).}\cite{Kamga2015_TPT}. Because pickup patterns vary in time, depending on the region, both taxi drivers and passengers suffer from long wait time according to \cite{KyungLee_2017TPT_Taylor, Kamga2015_TPT}. Dexterous taxi drivers could be benefiting from the knowledge they acquire about different locations demand patterns in different times e.g. by reacting to the increased demand due to the rain \cite{farber2015_QJE, Haggag2017AEJ}. However, the actual demand is unknown. In one hand taxi passengers experience long-wait time for vacant taxis in rainy conditions and on the other hand subway's humidity and temperature makes the transit inconvenient mode of transportation during summers in NYC. The characteristics of weather are the key to unlock the taxi demand problem in different weather conditions. Thus, any human mobility studies should model the impact of weather on individual agents.\\}\par 
\setlength\parindent{24pt}{Prolonged wait time worsened conditions more for passengers and less severe for drivers if we consider the weather in the taxi-passenger ridership problem. In a extreme atmospheric condition such as thunderstorm, prolonged wait time, causes damage to people in less secured areas against the weather in some regions. Even in the taxi-dominant areas of NYC, some people have experienced unfavorably long wait time on rainy days. Operation costs for taxis increase because of limited vision to spot hails, confusion in adapting to the situation, and increase in accidents in bad weather. However, cab-drivers were benefited from the abundance of customers in rainy samples in 2011 \cite{Kamga2015_TPT}. In essence, cab-drivers are daily income targeting \cite{Abel_brodeur_empirical_2018}. When cabbies met a certain cut-off because of the increased demand due to the weather, they called the day \cite{kamga2013_TRB92}. 
	\setlength\parindent{24pt}{ Waiting/cruising are among some strategies used by taxi drivers to approach passengers while the efficiency of these strategies was unknown for the researchers under different weather conditions. The historical peak-up data are satisfied demand by taxicabs. If the taxi does not present at an area, there would be no pickup record. In fact the primary object of current research is to incorporate supply to the demand problem.\\}\par
	Inferring demand is important for taxi drivers and policy-makers to quickly evaluate the accessibility and reliability of this transportation medium during worse weather conditions. Predicting the demand is difficult, because many factors are involving in human decision-making e.g. the number of available vacant taxis for street hail passengers. Therefore, pick-ups count, which is an indicator of the profitability of taxi companies in term of productivity has been used in lieu of the demand in most studies \cite{Zhang_2017_IEEETrBigData}. Despite the previous works, in this study, we considered mismatch indicators to infer the demand from partially observed data (pickups) using analysis of variance (ANOVA) and linear regressions.\\}\par

\section{Data Set}\label{sec3}
\setlength\parindent{24pt}{In 2013, 52800 licensed drivers with 13437 medallions served residents and travelers in major metropolitan areas like Manhattan, Brooklyn, etc. in NYC \cite{TLC2013}. It is estimated that between 400000 to 500000 daily trip hosted by medallion taxis during a study span from December 2008 to January 2010 \cite{Zhan2016}. Among many regions $95\%$ of pick-ups occur in Manhattan below the 96th Street and at JFK and LaGuardia airports {\cite{TLC2013}}. For instance, see the hotspots of NYC in Fig. \ref{fig:HotspotsNYC}.
	
	\begin{figure}
		\centering
		\adjincludegraphics[scale=.20,height=5cm,width=7cm]{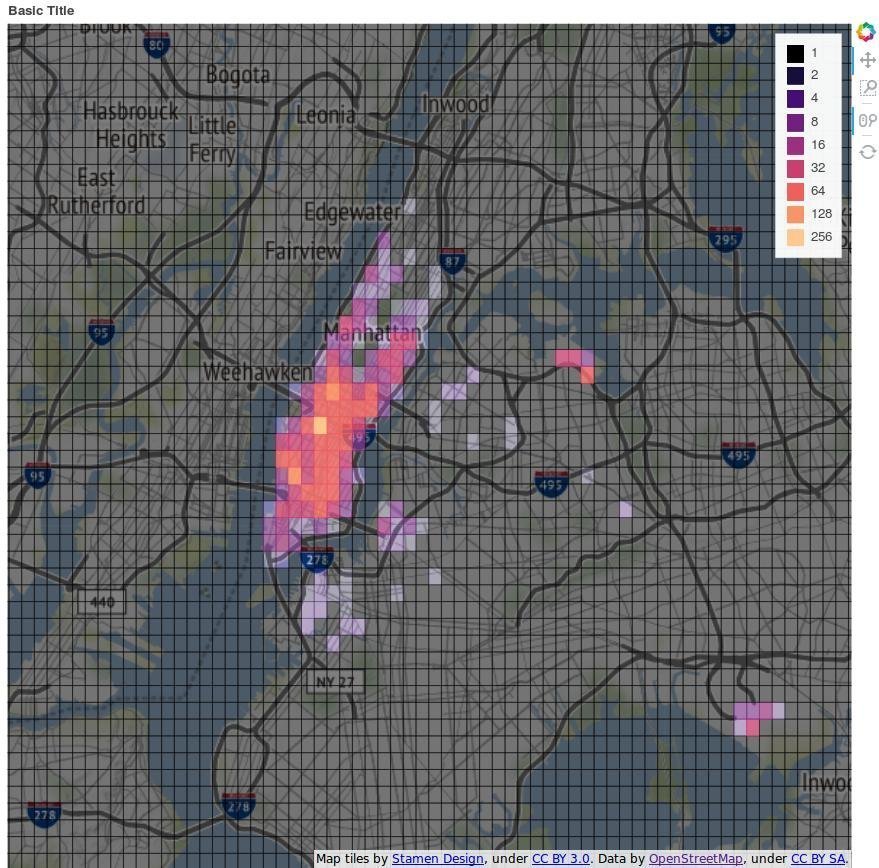}
		\caption{Density Plot for Number of Pickups in NYC in a Point of Time.} 

		\label{fig:HotspotsNYC}
	\end{figure}
}\par
\setlength\parindent{24pt}{\setlength\parindent{24pt}{One of the sensors typically embedded in every taxis in NYC is the global positioning system (GPS) sensor. The yellow taxis are equipped with built-in GPS report succinct information about this mode of transportation to a government agency in NYC. Taxi and Limousine Company (TLC) partially regulates the yellow taxi market in NYC and releases the GPS data to public semiannual. GPS sensors in different vehicles produce a tremendous amount of the coordinate data every second. Taxis are equipped with a GPS report accurate and traceable data about the location and time they were operating to the agencies, however in acquired data set only origin and destination of satisfied demand (pickups) were recorded. This data set could result in a better understanding of the transportation demand if they had been ingested with weather data. In human mobility domain, integration of GPS taxi data with exogenous weather information might result in improvements in taxi level of service, taxi market profitability, and customer satisfaction {\cite{HofmannITSC2005}}.\\}\par
	\setlength\parindent{24pt}{To this end, we streamed NOAA hourly weather information from three ground stations with a sampling rate of one hour are located in Central Park in Manhattan, La Guardia and JFK airports see Fig. \ref{fig:NCDC_stations}. The main surface weather stations are co-located where the majority of pick-ups/drop-offs occur (Central Park in Manhattan). The missing data of rain for one station is artificially generated using two other station data in the test set (detailed background is provided in \cite{KIM_Pachepsky_Hydrology2010305}).\\}\par
	\begin{figure}
		\centering
		\adjincludegraphics[scale=.3,height=5cm,width=7cm]{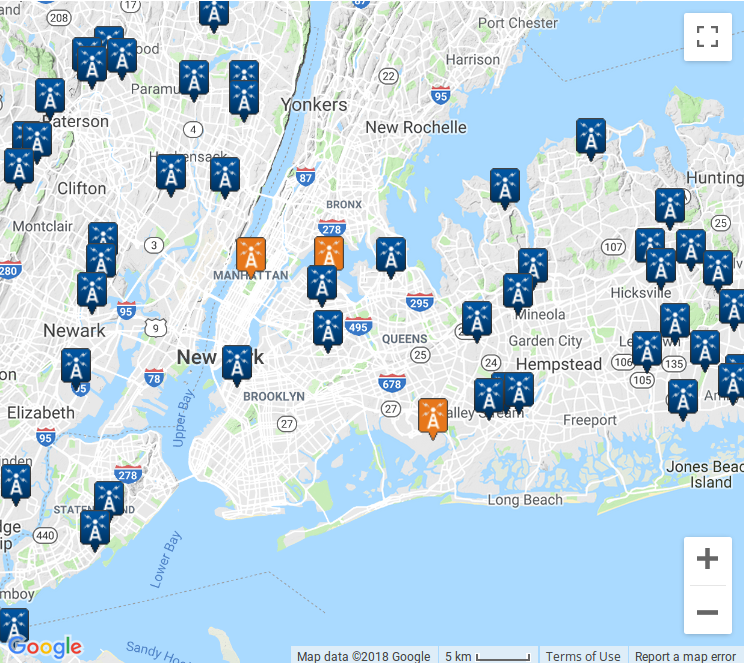}
		\caption{Data Obtained from Three Grounded Weather Stations.}
		\label{fig:NCDC_stations}
	\end{figure}
	\setlength\parindent{24pt}{Taxi data of NYC has a high spatiotemporal resolution. Annually, around 13000 medallions reached approximately 173 million trips (2013). Detailed ridership data of medallions was available online. Every recorded data pinpoints origin-destination (OD) pairs of when and where every individual ridership of yellow taxi took place rather a complete trajectory. However, the whole trip information is compact for example 2013 year the dataset size is roughly 45GB. The whole trip information is represented into key features (refer to Fig. \ref{fig:DataProcessingChartStac}). In the table of medallion dataset provided by TLC, there are 21 fields including medallion, hack license, pickup latitude, pickup longitude, dropoff latitude, dropoff longitude, pickup time, etc.\\}\par  
	\setlength\parindent{24pt}{The taxi supply field did not exist in data of taxis. The pickup record in the dataset of taxis is satisfied demand. The taxi supply in different locations at different times to associate the pickups with taxis. Because we had access to pickup and drop off location records at a given time we were able to calculate how many taxis operate at the city in sum. The major contribution of the current study is to generate individual taxi driver shift time sequence synthetically. Following two layers in Fig. \ref{fig:DataProcessingChartStac}, we have generated global supply. As a rule of thumb, the set of consecutive rides that have no more than 8-hour gap constitute a shift in Fig \ref{fig:Driver_shift_Visualization}.\\}\par
	\begin{figure}
		\centering
		\adjincludegraphics[height=5cm,trim=0 0 0 {.040\height},clip]{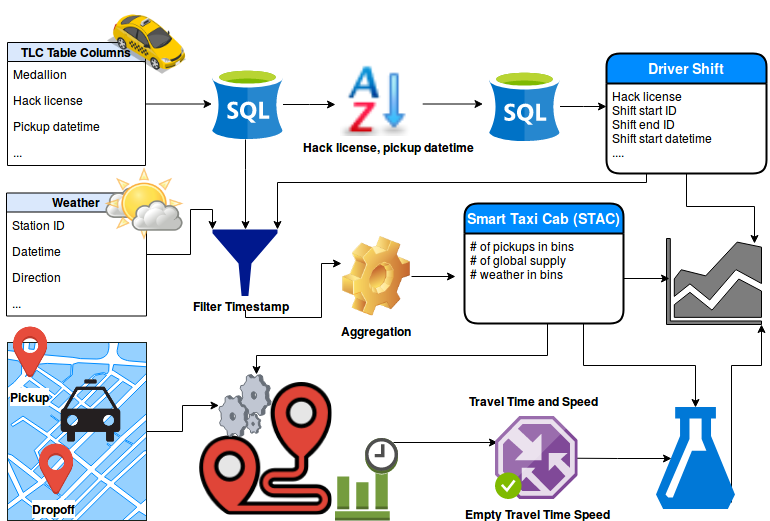}
		\caption{Data Processing Flowchart.}
		\label{fig:DataProcessingChartStac}
	\end{figure}
	
	\begin{figure}
		\centering
		\adjincludegraphics[scale=.4,trim={{.04\width} {.65\height} {.070\width} {.15\height}},clip]{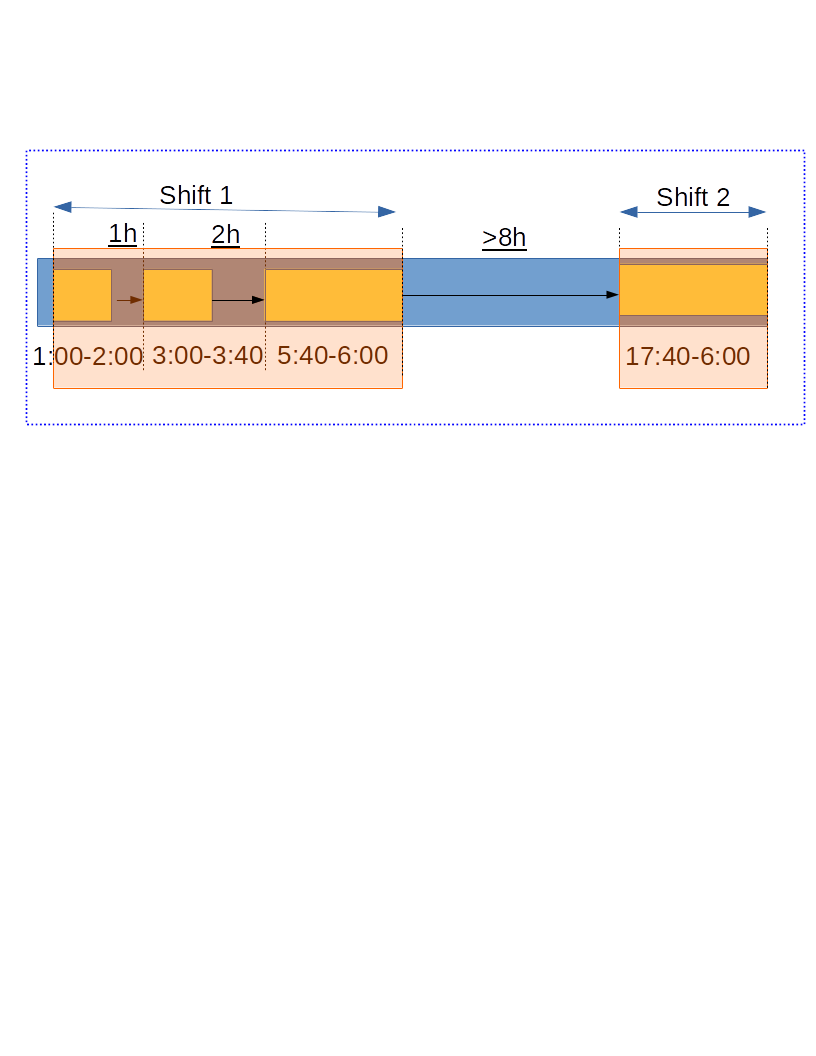}
		\caption{Depiction of Ingredients of a Shift.}
		\label{fig:Driver_shift_Visualization}
	\end{figure}
	
	\begin{figure}
		
		\hspace{.75cm}{\adjincludegraphics[height=5cm,trim={0 0 0 {.113\height}},clip]{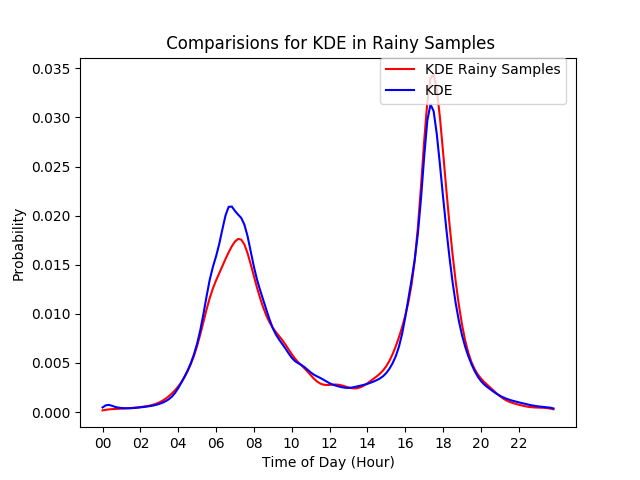}}
		\caption{Density Estimation for Shift Start Time in Day.}
		\label{fig:Driver_shift_start_Visualization}
	\end{figure}
	
	\begin{figure}
		\centering
		\adjincludegraphics[height=5cm,trim={0 0 0 {.113\height}},clip]{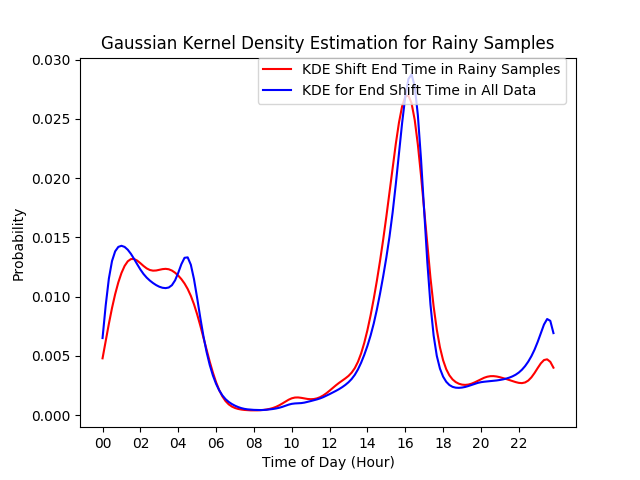}
		\caption{Density Estimation for Shift End Time in Day.}
		\label{fig:Driver_shift_End_Visualization}
	\end{figure}
	
	\setlength\parindent{24pt}{Anomalous shifts are easily tractable by an agent e.g. a driver rode for 30 day However, these anomalies are minutes. From Fig. \ref{fig:Driver_shift_start_Visualization} many Drivers start at one of two shift of morning and evening. The supply modeling using artificially generated shifts of the driver is generative modeling which generate whole supply with understanding only the distribution of the taxi shift start (PDF plot is a summation of several underlying Gaussian distributions). There were limitations regarding the definition of supply. On top of that taxi, supply is an approximate usually underestimated true supply, because we don't know how much in advance a driver starts his shift before making a fair. Alternatively, it is unclear when they decided to call the day. For when the taxi drivers have end taxis refer to Fig. \ref{fig:Driver_shift_End_Visualization}.
		Average travel time is 13.533 minutes and average empty ride time is 11.80 minutes in NYC in 2013. Detailed analysis of travel distance recorded in the trip distance is obtained with taxi meter readings refer to Fig. \ref{fig:Distancecompare}. As a brief conclusion in some location taxis spent more than three times of travel time empty. Because the trajectory is unknown for every taxis and the time spent on cursing/waiting, we resort to speed calculation for finding the localized supply. More detailed analysis of speed calculation is given in the forthcoming section.
		\begin{figure}
			\vspace{1cm}{
				\centering
				\vspace{-.4cm}{\adjincludegraphics[height=5cm,trim={0 0 0 {.07\height}},clip]{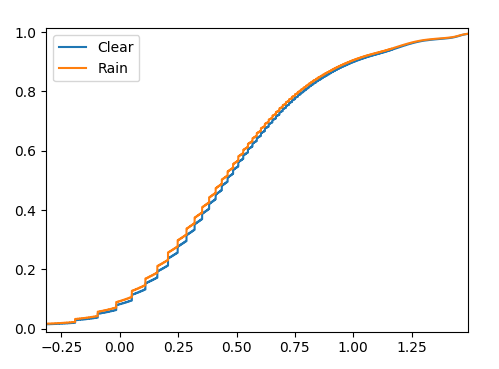}}
				\caption{CDF of Travel Distance (Log10 km).}
				\label{fig:Distancecompare}}
		\end{figure}
		
		\setlength\parindent{24pt}{The impact of weather on taxi-pick-ups depends on time. An hour precision is considered for time. We considered daytime, day of week, weekdays, weekend, day of month, month of year. \cite{kamga2013_TRB92,singhalTRPA2014,Xu2017IEEEITS2017,Kamga2015_TPT}. For statistically validating the experiments, only morning rush hours from 6- 10 A.M. and evening rush hours of 4- 8 P.M. are used.\\}\par

\section{Analysis}\label{sec4}
\setlength\parindent{24pt}{Time of day is a crucial factor. Data analysis is only provided for the morning peak period and evening peak period. Moreover, data exploration includes off-peak. The morning peak period identified 6-10 a.m. according to The Port Authority of NY and NJ \cite{Ukkusuri2016EURO}. Evening peak hours are from 4-8 p.m. \cite{NJ_NYPort}.
	Despite a low pickup for taxis exists on weekdays in some locations like Avenue 6th in Manhattan always, always there is more demand for taxis in weekdays morning peak hours similar to \cite{FerreiraIEEEVisual2013}. The reason for low pickup is congestion during the weekdays.  Fewer pickups of taxi occur in weekdays in regions like Avenue C in Manhattan. However, as shown in Fig.\ref{fig:pick_up_347314_ave_c_nyc}, in Avenue C more pick-ups occur on weekends.\\}\par
\begin{figure}
	\centering
	\adjincludegraphics[height=5cm,trim={0 0 0 {.093\height}},clip]{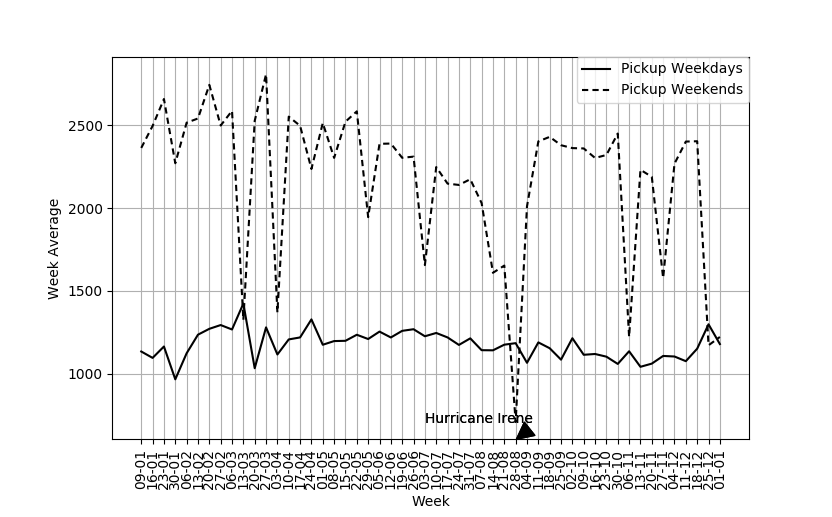}
	\caption{Difference in Weekday and Weekend Pickup Counts in 2011.}
	\label{fig:pick_up_347314_ave_c_nyc}
\end{figure}
\begin{figure}
	\adjincludegraphics[height=5cm,trim={0 0 0 {.073\height}},clip]{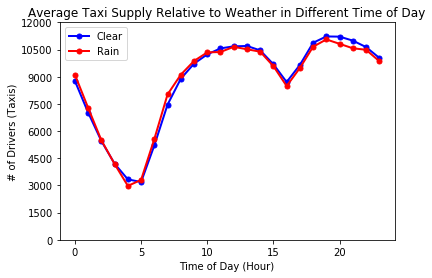}
	\caption{Average Taxi Supply Relative to Weather in Different Time of Day.}
	\label{fig:supplylevelTaxiNYC}
\end{figure}
\setlength\parindent{24pt}{Effect of rainy weather on weekdays is different than the weekends. The effect of rain on morning peak periods in weekends results into less increase in the number of pickups and less increase of mismatches for demand. Yet, in rainy weekday morning peaks more demand exists (higher number of pickups and higher pickups per driver). In the evening peak, the highest number of pickups per driver and number of pickups have been observed while the magnitude of mismatch is unclear from the plots of weekdays in citywide. 
	\begin{figure}
		\centering
		\adjincludegraphics[height=5cm,trim={0 0 0 {.06\height}},clip]{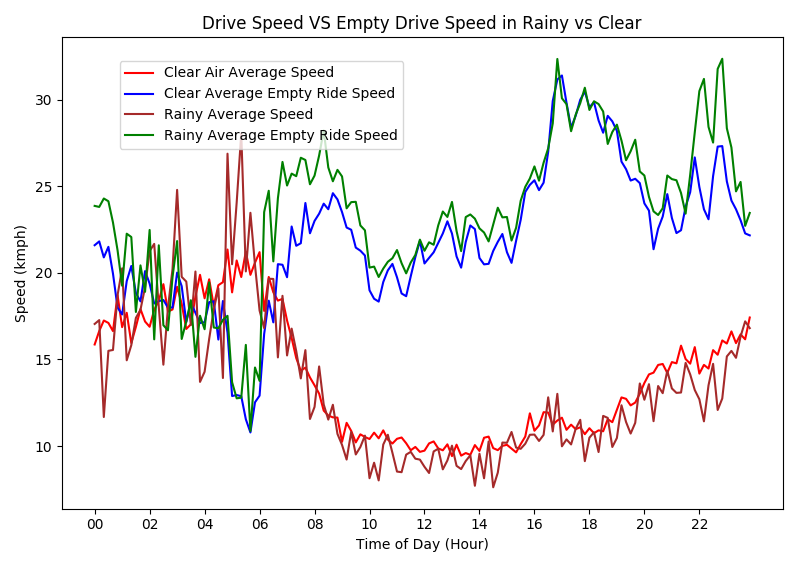}
		\caption{How Fast Taxis Found their Fairs in a Location.}
		\label{fig:PickupswithhigherResolution}
	\end{figure}
	We looked closely to empty driving time for location in Manhattan to examine traffic condition was measured by space mean speed, these. In the plots of traffic condition, empty travel speed means how quickly an empty taxi get its fair. The data between 2 am and 6 am are noisy (brown) because relatively fewer number of pickup data were available comparison to peak periods \ref{fig:PickupswithhigherResolution}.
	\begin{figure}
		\centering
		\vspace{2.18cm}{\adjincludegraphics[trim=0 1cm 0 8cm,scale=.35]{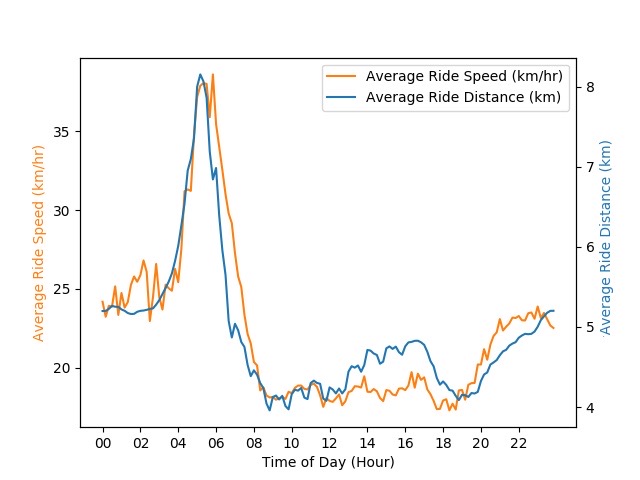}}
		\caption{Travel Speed in Different Time in Day in NYC.}
		\label{fig:my_label}
	\end{figure}
	To compare and contrast the result of macroscopic with speed of hot spot, citywide travel speed provided in Fig. \ref{fig:PickupswithhigherResolution}. The shortest trips are observed at 6 p.m. In weekends, taxis operate with more freedom because of less traffic. The other factor that explains the increase in the taxi pick-ups is the unavailability of substitute transportation during off-peak hours especially at midnights when unavailability of alternatives like car-renting, or less availability of transit ridership, and its limited schedule. The weather shock has made more mismatch into a specific location. The higher profit time for pickups for the taxi is late at night, in midnight and early in the day. However, transit ridership has peaked at evening when people come back at work and later in the mornings when people go to work.
	\begin{figure}
		\centering
		\adjincludegraphics[height=5cm,trim={0 0 0 {.073\height}},clip]{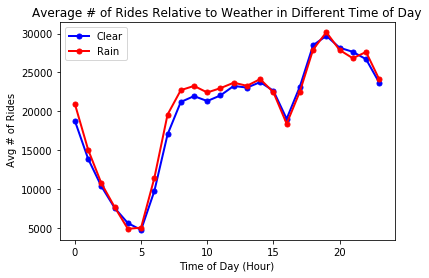}
		\caption{Average of Taxi Pickups in Clear Weather versus Rainy.}
		\label{fig:my_label}
	\end{figure}
	\begin{figure}
		\centering
		\adjincludegraphics[height=5cm,trim={0 0 0 {.073\height}},clip]{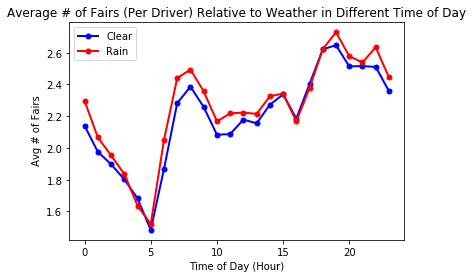}
		\caption{Average of Number of Pickups Per Driver in Clear Weather versus Rainy Weather.}
		\label{fig:my_label}
	\end{figure}
	
	\begin{figure}
		\vspace{-.5cm}{
			\centering
			\adjincludegraphics[height=5cm,width=10cm,trim={1.5cm .3cm 0 {.073\height}},clip,scale=.5]{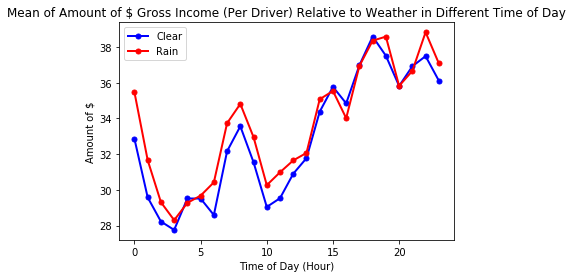}}
		\caption{Average Income of Taxi Drivers in Clear Weather versus Rainy Weather.}
		\label{fig:AvgIncomeTaxiDrivers}
	\end{figure}
	\setlength\parindent{24pt}{To validate our result, we used nonparametric test of Mann-Whitney which is equivalent to parametric {ttest} and a nonparametric test for median using permutation analysis. Finally, we cross-validate our method on simulation data. The choice for an opting nonparametric test instead of well-used analysis of variance test (ANOVA) is because of the violation of normality underlings of data and hetroscdastisity assumptions. We used a test for comparisons in means and medians of two samples, obtained by permutation analysis as well as on the original data set. The chose of permutation analysis is due to confounding factor time (special days). To remove the time component we randomly made dates with 4 morning hours in the first experiment. In the second experiment, we repeat the same procedure for evening peak hours. A once paired sample test of Wilcoxon signed rank test (the equivalent of Mann-Whitney test) obtained null hypothesis of equality in means of two independent samples with a dependent variable of the average number of fairs per drivers in an hour in rain versus clear conditions was performed. Kruskal-Wallis test used for comparisons of medians of two samples. The statistics of analyzing results are provided in section \ref{sec14} (Appendix Tables \ref{tab1} and \ref{tab2}).\\}\par

%
	

\section{Conclusions and Future Work}\label{sec5}
\setlength\parindent{24pt}{It is cited that rainy weekdays increase the demand for businesses and decreases discretionary trips. In contrast, during the weekends, an increase in precipitation reduces recreational needs like visiting the museums for different transportation mediums \cite{Sabir2011Dissertation, Guo2007}. We obtained the different result for morning peak hours that during that period if the higher leisure activities associated with weekends than weekdays, higher demand were observed due to the rain in weekends for these trip purposes.\\}\par 
\setlength\parindent{24pt}{The effect of weather depends on location and the mismatch factors. Mismatch variations are higher due to weather in a hotspot as observed in Fig \ref{fig:PickupswithhigherResolution}. In that location, the drivers who were present at that location found the fairs easily during the evening rain even in the worst traffic condition. In result, when the higher pickup associated with location and time, a higher magnitude of demand and mismatch exists. The result of this experiment identified that higher demand exists in a citywide during the rain as well. We illustrated that the precipitation income is a better add-in to taxi operation income driven factors, but the higher chance of crash associated with higher precipitation. However, adding other weather-related factors such as the temperature to baseline remained unexplored. The taxi operation in some degree depends on external factors such as special events that we did not consider. In fact, we eliminated all time confounding factors using permutation analysis and the implications were discussed.\\}\par
\setlength\parindent{24pt}{The projected evening demand was less likely to be affected by rain a macroscopic perspective. However, from data of specific location, it is highly affected in evening peak hours. When the taxis could not fluid from one point to another, in dense areas, more short trips are results of more demand when the demand is higher. We did not cross-validate our analysis with external data sources such as traffic. However, analysis compared to TLC hourly income from the financial perspective compared to calculated with partially observed data as a process to result in prosperity \cite{oleyaeimotlagh2014Strategy}. Taxi drivers decision to leave the shift early seem an irrational decision as income follows super graph in every time of day relative to the base model. Comparison of time of day income they topically obtain higher income while each trip adds-up less to total income. The taxis shift appears in our data when there was a pickup record. More tuning the shift start and end time and decision of driver is recommended for future studies. \\}\par

\section{Acknowledgments}\label{sec6}
The authors have been supported by the High Performance Computing (HPC) team at Institute for Simulation and Training at University of Central Florida .
\bibliography{IET-Submission-STAC}

\vfill\pagebreak
\section{Appendices}\label{sec14}
The objective of statistical analysis in the Appendix is to present
statistical methods applied in complement to data, which were illustrated in the  analysis section of current article (see Table ~\ref{tab1} and  Table ~\ref{tab2}).
\vfill\pagebreak
\vspace{19cm}
\begin{table}[!h]
\fwprocesstable{Result of Nonparametric Statistical Analysis for Morning Peak Hours in the Different Days of Week.\label{tab1}}
{\begin{tabular*}{\textwidth}{@{\extracolsep{\fill}}lccccc}\toprule
		& &Permutation  &Permutation  &Observation &Observation\\
		Day of Week &Test  &Statistics &pvalue   &Statistics &pvalue\\
		\midrule
		Weekday &    Wilcoxon &389471837.0 &    0.0 &- &- \\
		Weekday &    Mann-Whitney &1203250590.5 &0.0 &3103.0 &0.00042 \\
		Weekday &    Kruskal &86491.40403&   0.0 & 11.14422 &0.00084 \\
		Weekend &    Wilcoxon &307599058.0 &   0.0 & -&- \\
		Weekend &    Mann-Whitney &3540319.0 &    0.00132727.0 &727.0 &0.45800 \\
		Weekend &    Kruskal &102428.47286 & 0.0 & 0.01207 &0.91250\\
		\botrule
\end{tabular*}}{}
\end{table}

\begin{table}[!h]
\fwprocesstable{Result of Nonparametric Statistical Analysis for Evening Peak Hours.\label{tab2}}
{\begin{tabular*}{\textwidth}{@{\extracolsep{\fill}}lcccc}\toprule
		&Permutation  &Permutation  &Observation &Observation\\
		Test  &Statistics &pvalue   &Statistics &pvalue\\
		\midrule
		Wilcoxon &1224426201.0 &0.0 &- &-\\
		Mann-Whitney &2848413284.5 & 0.0 &3103.0 &0.00042\\
		Kruskal &27775.81349 &0.0 &0.01642 &0.89800\\
		\botrule
\end{tabular*}}{}
\end{table}

\end{document}